# Optomechanical Properties of Stretched Polymer Dispersed Liquid Crystal Films for Scattering Polarizer Applications

Ichiro Amimori[†], Nikolai V. Priezjev[*], Robert A. Pelcovits[*], Gregory P. Crawford[†, *]

[†]Division of Engineering, Brown University, RI 02912

[*]Department of Physics, Brown University, Providence, RI 02912

## Abstract

A scattering polarizer is created by subjecting a polymer dispersed liquid crystal (PDLC) film to tensile strain. The optomechanical properties of the film are investigated by simultaneously measuring the stress-strain and polarization dependent optical transmission characteristics. The correlation between transmittances of two orthogonal polarizations and the stress-strain curve reveals that the polymer orientation as well as the droplet shape anisotropy influences the liquid crystal alignment within the droplets. A Monte-Carlo simulation based on the Lebwohl-Lasher model is used to explain the subtle influence of polymer orientation on liquid crystal alignment.

I.     **Introduction**

Power consumption in liquid crystal displays (LCD) is of the utmost importance for portable applications. In typical LCDs utilized in laptop computer applications, the optical efficiency is approximately 6%.[1] One of the primary reasons for the poor efficiency of an LCD is the presence of absorbing polarizers. Conventional polarizers based on absorption are iodine doped polyvinyl alcohol (PVA) films that absorb more than 50% of the incident light (see Figure 1(a)). In order to improve the optical throughput and efficiency of LCDs, a light recycling technique has been developed which uses non-absorbing polarizers.[2] The ideal concept of a non-absorbing polarizer is one which efficiently reflects or backscatters light rather than absorbing it. With a light recycling mechanism, the rejected light can be redeployed if the polarization can be manipulated (see Figure 1(b) and (c)).

There are several approaches to non-absorbing polarizers, all of which fall into two basic categories: reflection and light scattering. In order to create a reflective polarizer, a periodic medium must be used, such as uniaxially stretched multilayered polymer thin films with alternating indices of refraction[3,4] and cholesteric liquid crystal layers with a quarter wave plate[5] as illustrated in Figure 1(b). In order to create a light scattering polarizer, a composite material must be used, such as uniaxially stretched phase separated polymer films,[6,7] isotropic particle dispersed uniaxially stretched polymer films[8], and uniaxially stretched polymer dispersed liquid crystal (PDLC) films.[9]

From a technology perspective, scattering polarizers are less expensive than reflective polarizers due to their simple structure and fabrication process. The polarization selectivity of scattering polarizers can be created by the liquid crystal alignment within the droplets. There have been many basic studies of liquid crystal alignment in spherical and elliptical droplets.[10-18] Aphonin and coworkers studied the optical property of a stretched PDLC film with bipolar nematic droplets.[9, 19] They described both the transmittance dependence on strain and the scattering profile based on the anomalous diffraction approximation.[20, 21] The liquid crystal alignment in a droplet is determined by the minimization of the droplet free energy with tangential boundary conditions at the droplet surfaces. Zhao and coworkers investigated the order parameter of a stretched PDLC film as a function of strain by means of infrared dichroism.[22] Within a two-dimensional model, Chan numerically studied the nematic director orientation in elliptical bipolar droplets using the Leslie-Ericksen and Frank continuum theories.[23, 24]

Although the effects of the polymer orientation surrounding the droplets was not considered in these papers, it is well known that the interaction of liquid crystal molecules at the polymer surface strongly influences the liquid crystal orientation within the droplet.[25] Polymer orientation was found to strongly influence liquid crystal molecules at the planar surfaces used in display applications.[26] Since polymer chains align during the PDLC film stretching process,[27] subtle changes in the liquid crystal alignment result from the polymer alignment induced by the tensile strain. In this paper we report the results of a study of the optomechanical properties of a PDLC film and demonstrate that both droplet shape anisotropy and polymer orientation play a role in the final nematic director configuration. We also employ Monte-Carlo simulations based on the Lebwohl-Lasher model[28, 29] to explain the effect of polymer orientation on liquid crystal alignment.

### II. Materials and Experimental Setup

PDLC films used in this study consisted of dispersions of the liquid crystal E7 (*EM Industries, Inc.*) and polyvinyl alcohol PVA205 (*Kuraray Co., Ltd.*), which is a water-soluble polymer. A 20 wt. % aqueous solution of PVA is mixed with liquid crystal and then dispersed in water.[10] We used an ultrasonic processor (CPX-400, 400 W, 20 kHz, *Cole-Parmer Instrument Co.*) at 40% output with 1/8" microtip to emulsify the liquid crystal into the PVA aqueous solution. The emulsion was then coated on a smooth polyethylene terephthalate (PET) substrate using a Meyer Bar and processed under ambient conditions. After the water evaporated, the polymer film was carefully peeled from the substrate. The film thickness was approximately 16±4 μm and the concentration of liquid crystal in the film was 25 wt. %.

Figure 2 shows scanning electron microscope (SEM) images of a PDLC film before and after stretching (100% strain). The director within the droplet aligns parallel to the stretch direction (the major axis of the ellipse) [15] thereby creating a refractive index difference between the axes parallel and perpendicular to the stretch axis. The refractive index of PVA is nearly isotropic since its birefringence induced from stretching is relatively small. As a result, a large refractive index mismatch exists in a direction parallel to the stretch axis, while a refractive index match occurs in a direction perpendicular to the stretch axis because the ordinary refractive index, $n_o$, of the liquid crystal is approximately equal to the index of the polymer, $n_p$, as illustrated in Figure 3. Therefore one polarization state in which the two indices of refraction are the same can transmit, while the orthogonal polarization state is scattered due to the refractive index mismatch. The aspect ratio of the stretched droplet shown in Figure 2(b) is approximately 4:1 for 100% strain. The discrepancy between the aspect ratio (4:1) and the strain (100%) is attributed to the width contraction of the film during stretching governed by Poisson's ratio,[27] which describes the contraction in the direction perpendicular to the tensile axis.

Figure 4 shows the experimental setup for the simultaneous measurements of the optical and mechanical properties of PDLC films. A He-Ne laser (633 nm) was used as the light source and directed at normal incidence on a dog-bone shaped sample through a rotating polarizer. Polarizations parallel and perpendicular to the tensile axis were investigated. Dog-bone shaped samples are typically used in stress-strain measurements to create a uniform stress-strain relationship in the specimen.[30] The sample was uniaxially stretched by a tensile tester Minimat 2000 (*Rheometric Science, Inc.*) with a draw rate of 0.1 mm/min. The stress of the sample was measured using a load cell and the intensity of light not scattered by the sample was simultaneously detected by a photo detector.

### III. Experimental Results and Discussions

The stress-strain curve in Figure 5(a) shows the transmittances of two polarization states, parallel ($T_{//}$) and perpendicular ($T_\perp$) to the tensile axis. In Figure 5(b), we show the transmittance difference, $\Delta T = T_\perp - T_{//}$, vs. strain. It can be seen that $\Delta T$ saturates only after a significant decay in the stress has occurred. Figure 5(c) shows an expanded view of Figure 5(b) within the strain range from 0 to 30%. The stress-strain curve is linear up to the maximum value of the stress (elastic region) and then begins to decrease (plastic region). In the elastic region, $\Delta T$ increases slightly and then more significantly as the plastic region is reached, as shown in Figure 5(c). The parameter $\Delta T$ reflects the structural change in the liquid crystal alignment along the tensile axis. The point where stress begins to decay in Figure 5(b) is called the yield point.[31] The decay of the stress suggests that the polymer matrix begins to disentangle due to the breaking of crosslinks. We believe that this polymer disentanglement, which decreases the friction coefficient of the polymer matrix, leads to the observed stress-strain behavior.[32] Based on these observations, we labeled the large stress decay regime after the yield stress as the disentanglement regime. The constant stress regime after the disentanglement regime is often described by the dashpot (slider) model.[30, 32] We labeled this constant stress regime due to the sliding of fully oriented polymer chains as the sliding regime. Given the results of our measurements, we propose that polymer chain alignment is another mechanism which influences the liquid crystal structure in addition to the droplet shape anisotropy.

In order to further study the polymer chain alignment, we investigated the pure PVA film. The experimental apparatus used to simultaneously measure birefringence and stress-strain is shown in Figure 6(a). To effectively measure the birefringence, the optic axis (tensile axis) was oriented at 45 º relative to crossed polarizers. Figure 6(b) shows the dependence of the birefringence and stress of the PVA film on strain. The transmittance is described by the following equation:[33]

$$T = \frac{1}{2}\sin^2\left(\frac{\pi}{\lambda}\Delta n(\varepsilon)d(\varepsilon)\right) = \frac{1}{2}\sin^2\left(\frac{\pi}{\lambda}\Delta n(\varepsilon)d_0[1-\alpha\varepsilon]\right), \qquad (1)$$

where $\lambda$ is the wavelength of incident light (633 nm), $\Delta n(\varepsilon)$ is the birefringence at strain $\varepsilon$, $d(\varepsilon)$ is the thickness at strain $\varepsilon$, $d_0$ is the initial thickness, and $\alpha$ is the contraction ratio of the film thickness during tensile strain. The film thickness was calculated by $2 \times 2$ matrix method[34] from measured transmission spectra in the range from 600 to 800 nm. The value of $\alpha$ was determined to be 0.492 from the slope of the thickness vs. strain relation in Figure 6(c). This value can be interpreted as Poisson's ratio, where the value of $\alpha = 0.492$ is close to 0.49 of natural rubber.[31] In the stress-strain curve of the PVA film shown in Figure 6(b), there are three regions, which are the elastic region, plastic region and intermediate region. The transition point at which the slope of $\Delta n$ vs. strain changes corresponds to the strain where the stress-strain curve enters the plastic region. This result provides additional evidence that the polymer chains start to align at this point.

### IV.    Monte-Carlo Simulations for Stretched PDLC

In order to further understand the effect of polymer alignment on the liquid crystal, we employed Monte-Carlo simulations of the Lebwohl-Lasher lattice model[28, 29] of nematics confined to elliptical cavities with homogeneous boundary conditions enforced by a surface anchoring potential.[35] Although we cannot separate out the effects of shape anisotropy and polymer alignment, both effects are included in the model. This model represents each mesogenic molecule as a three-dimensional spin vector. These spins are free to rotate about their centers, which are fixed on the lattice sites.

The bulk Lebwohl-Lasher pair potential describing the interaction of mesogens within the droplets is given by:

$$U_{ij} = -e_b P_2(\cos\theta_{ij}), \qquad (2)$$

where $\theta_{ij}$ is the angle between two spins located at lattice sites $i$ and $j$, $P_2$ is a second rank Legendre polynomial, and $e_b$ is a positive constant for nearest-neighbor sites and zero everywhere else.

The homogeneous surface anchoring is introduced via a group of boundary layer spins located on the outside surface of the droplet wall. Here we use the boundary layer spins to model the polymer surface. The interaction of these spins with those inside the droplet wall is described by the potential

$$U_{ik} = -e_s P_2(\cos\theta_{ik}), \qquad (3)$$

where spin $i$ is located inside the boundary (representing a group of liquid crystal molecules) and spin $k$ is the nearest-neighbor spin that belongs to the boundary layer (representing polymer molecules). The constant $e_s$ is the surface anchoring strength. The total potential energy in a droplet is given by the sum of these potentials over all pairs of molecules. To introduce a preferred azimuthal anchoring direction, the molecules in the boundary layer were assumed to be oriented permanently along the meridians of the ellipsoid, to mimic the interaction with the ordered polymer chains.

Figure 7(a) and (b) show the orientation of the spins within the droplets with $e_s = 0$ and $e_s = e_b$, respectively. In Figure 7(b), the existence of the boundary layer spins outside the surface of the droplet wall can be seen. The size of droplet is 40 spins along the major axis and 30 spins along each of the two minor axes of ellipse. While there is no significant pictorial difference between but the two figures, we found that the nematic order parameter in the presence of polymer alignment is 0.63, compared to 0.60 in the absence of alignment, thus indicating that the polymer does indeed influence the liquid crystal orientation.

## V.     Conclusions

By simultaneously measuring the stress-strain characteristics and the transmittance parallel and perpendicular to the stretch axis of PDLC films, we determined the relationship between the polarization properties and stress-strain properties of PDLC scattering polarizers. We found that in addition to droplet shape anisotropy the polymer orientation during stretching also contributes to the liquid crystal alignment within the droplets. We also studied the director configuration numerically using Monte-Carlo simulations of the Lebwohl-Lasher model where both shape anisotropy and surface anchoring are included. The results of this simulation indicate that the polymer alignment at the interface increases the droplet order parameter along stretch axis, consistent with the results of the optical and mechanical experiments.

## VI. Acknowledgements

Financial support was provided by *Fuji Photo Film Co. Ltd.* (I.A.) and the National Science Foundation under Grants. Nos. DMR-9873849 and DMR-0131573 (N.P. and R.A.P), and DMR-0079964 and DMR-9875427 (G.P.C). Helpful discussions with Professor R. Clifton are gratefully acknowledged.

**Captions**

Figure 1  Configurations of various polarizers used in LCDs: absorbing (a), reflective (b), and light scattering (c).

Figure 2  SEM images of E7/PVA PDLC films before (a) and after (b) stretching.  The strain is 100% after stretching (b).

Figure 3  The distinction in refractive indices between liquid crystal droplets and PVA in stretched PDLC films and the index matching condition, $n_p = n_o$, of a stretched PDLC film.

Figure 4  Experimental setup for the simultaneous measurement of optical and mechanical properties of stretched PDLC films.

Figure 5  The stress-strain curve (solid line) and the transmittance properties (data points) of light polarized parallel, $T_{//}$, and perpendicular, $T_\perp$, to the stretch axis of the PDLC film are shown in (a).  The transmittance difference $\Delta T$ vs. strain and the stress-strain curve is shown in (b).  The expanded $\Delta T$ vs. strain is shown in (c).  The solid and dashed circles in the insets indicate crosslinks and disentanglements of polymer chains, respectively.

Figure 6  Experimental setup for the simultaneous measurement of optical and mechanical properties for analyzing the birefringence property of the PVA film is presented in (a).  The stress-strain curve and the birefringence of PVA film during stretching are shown in (b). The $d\Delta n/d\varepsilon$ curve is overlaid in (b) to show the slope change of the $\Delta n(\varepsilon)$ curve. The dependence of PVA film thickness on strain is shown in (c).

Figure 7  Liquid crystal alignment in elliptical PDLC droplets simulated by the Lebwohl-Lasher model: (a) without polymer alignment and (b) with polymer alignment. The solid circles indicate the droplet wall and the spins outside the droplet wall in (b) indicate the inclusion of boundary layer spins.

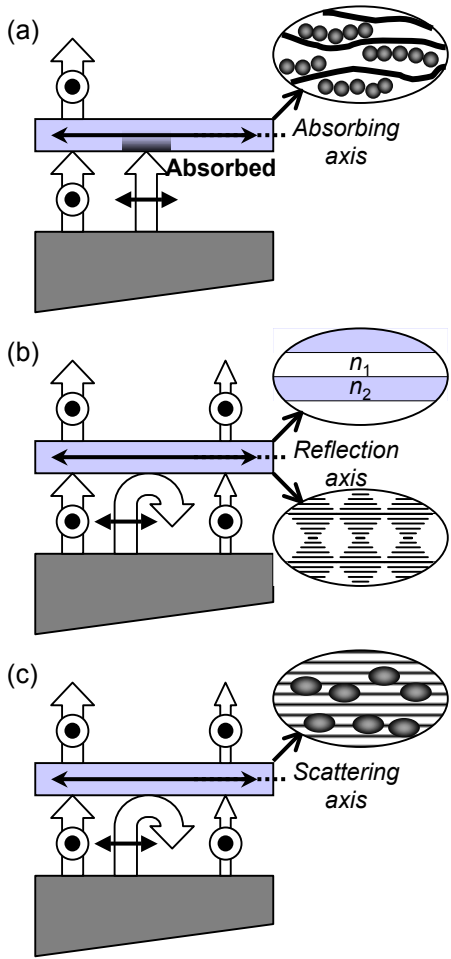

Figure 1
I. Amimori, *et al.*, Optical and Mechanical Properties of Stretched PDLC Films for Scattering Polarizers

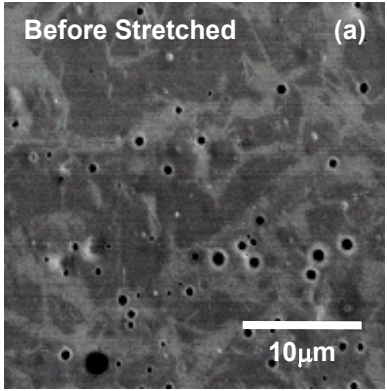

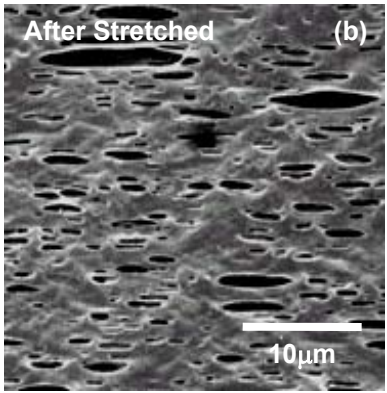

Figure 2
I. Amimori, *et al.*, Optical and Mechanical Properties of Stretched PDLC Films for Scattering Polarizers

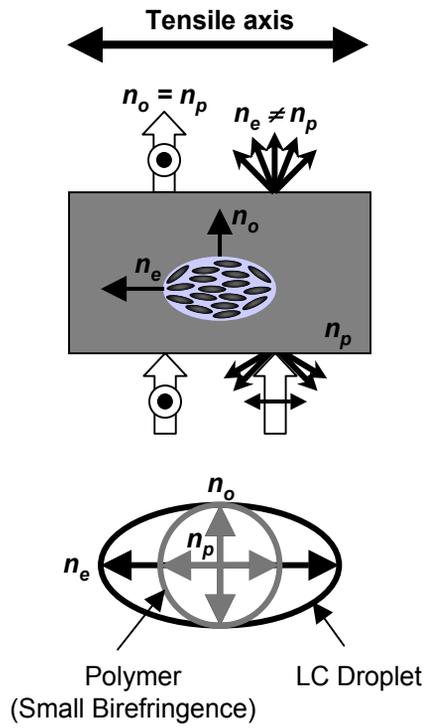

Figure 3
I. Amimori, *et al.*, Optical and Mechanical Properties of Stretched PDLC Films
for Scattering Polarizers

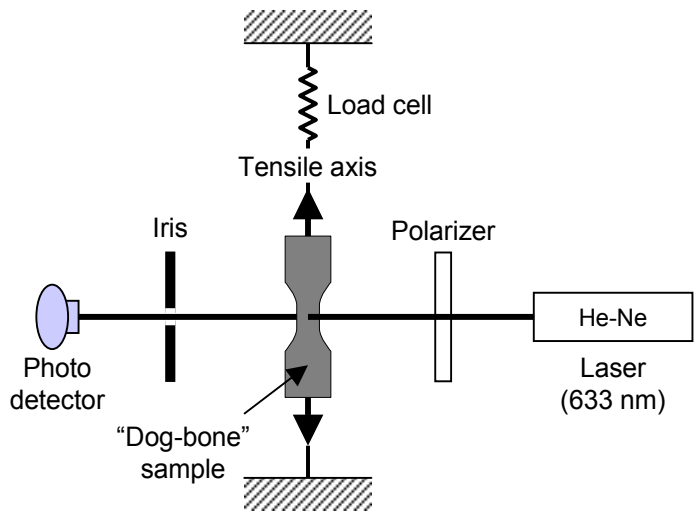

Figure 4
I. Amimori, *et al.*, Optical and Mechanical Properties of Stretched PDLC Films for Scattering Polarizers

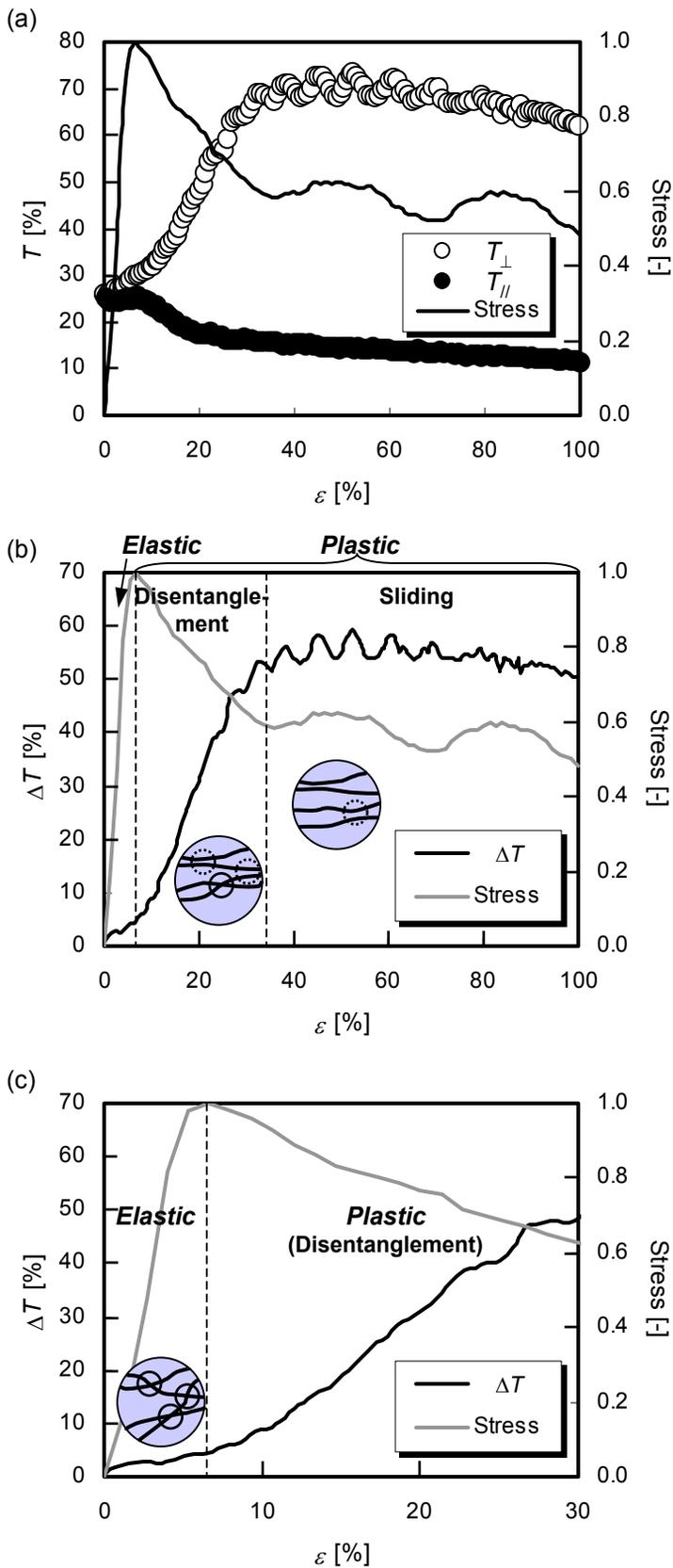

Figure 5
I. Amimori, *et al.*, Optical and Mechanical Properties of Stretched PDLC Films for Scattering Polarizers

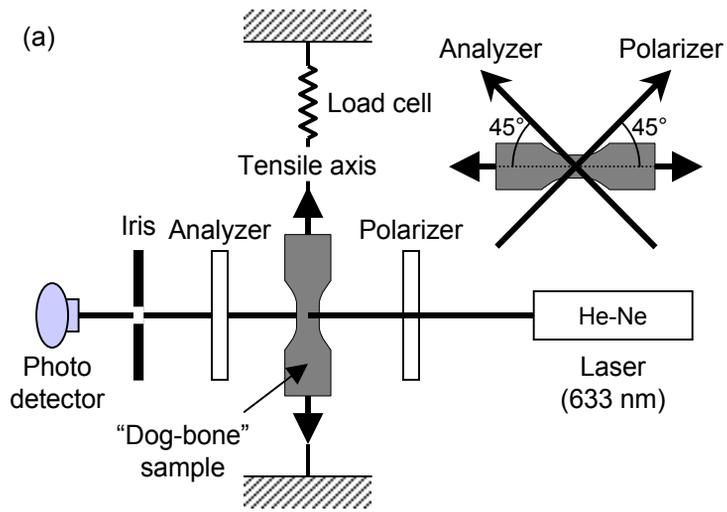

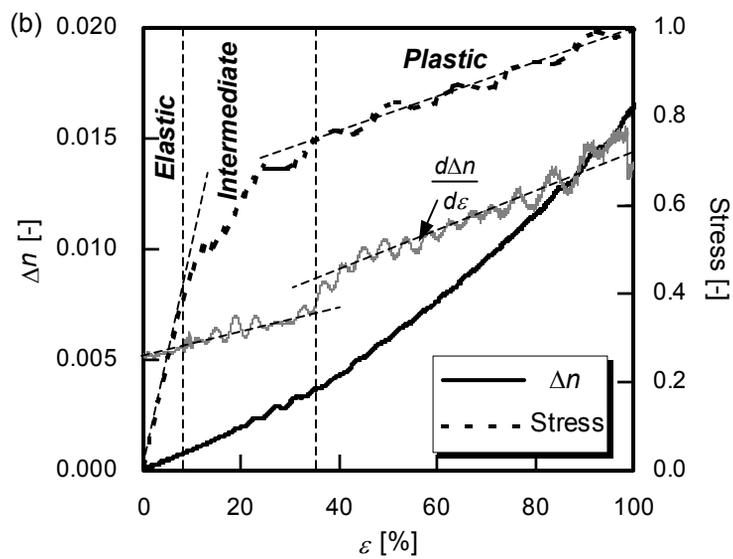

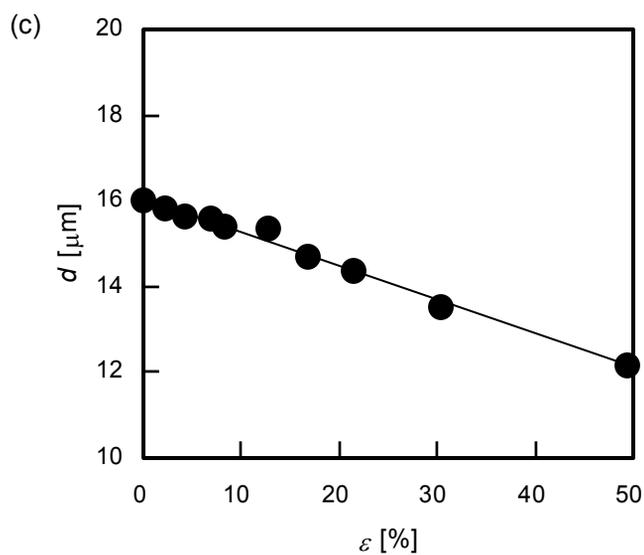

Figure 6
I. Amimori, *et al.*, Optical and Mechanical Properties of Stretched PDLC Films for Scattering Polarizers

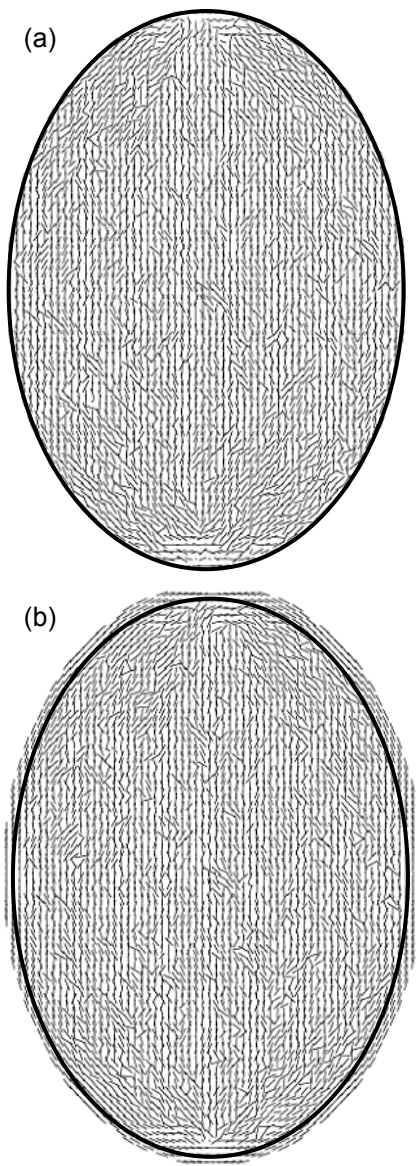

Figure 7
I. Amimori, *et al.*, Optical and Mechanical Properties of Stretched PDLC Films for Scattering Polarizers